\documentclass{iau}
\usepackage{graphicx} 

\title[IAUS291.~~Timing noise and pulse profiles] 
{Timing noise and the long-term stability of pulsar profiles} 

\author[A. Lyne]  
{Andrew Lyne}

\affiliation{Jodrell Bank Centre for Astrophysics, School of
Physics and Astronomy,\\ University of Manchester, Manchester M13 9PL,
UK \\ email: {\tt agl@jb.man.ac.uk}} 

\pubyear{2012}
\volume{291}  
\jname{\mbox{Neutron Stars and Pulsars: Challenges and Opportunities after 80 years}}
\editors{J. van Leeuwen, ed.} 
\begin{document}

\maketitle

\begin{abstract}
It has recently been shown that there is a close correlation between
the slowdown rates and the pulse shapes of six pulsars, and between
the slowdown rates and the flux density of three others.  This
indicates that these phenomena are related by changes in the current
flows in the pulsar magnetospheres.  In this paper we review the
observational status of these studies, which have now been extended to
a total of 16 pulsars having correlated slowdown and pulse emission
properties.  The changes seem to be due to sudden switching between
just two discrete magnetospheric states in the well-known processes of
mode-changing and pulse nulling.  We also address how widespread these
phenomena are in the wider pulsar population.

\end{abstract}


\firstsection 
\section{Introduction}
Timing noise is low-frequency fluctuation in the rotation-rate of
pulsars and is evident in the timing residuals of all young and
middle-aged pulsars. The basic properties of timing noise were reviewed
recently by \cite[Hobbs et al. (2010)]{hlk10} who presented the results of an
analysis of the rotation of 366 pulsars.  In summary, timing noise is
seen as smooth changes in the timing residuals (and rotation
frequency). The timing residuals are often asymmetric, with peaks and
troughs having different radii of curvature; the variations are
often quasi-periodic with timescales which are typically 1-10
years. For long, it was thought to arise from the fluid interiors of
the neutron stars.

A breakthrough was made in 2006, when \cite[Kramer et al.]{klo+06}
showed that the timing noise in the long-term intermittent pulsar
B1931+24 could be explained in detail by switching in the magnitude of
the slowdown rate $\dot{\nu}$ betwen the "ON" and "OFF" emission
states of the pulsar, indicating that changes in the current flow from
the pulsar resulted in changes of both the radio emission and of the
braking torque.  The implication that changes in magnetospheric
currents could alter pulsar emission properties as well as slowdown
rate led \cite[Lyne et al. (2010)]{lhk+10} (hereafter LHKSS) to study
the detailed pulse profiles of some of those pulsars having the
largest amounts of timing noise.  They demonstrated that six of these
pulsars exhibited the well-known phenomenon of mode-changing, in which
a pulsar switches abruptly between two stable profiles.  Moreover, in
all six pulsars, there was a high degree of correlation between the
pulse shape and slowdown rate, the pulsars switching rapidly between
low- and high-spindown rates.

Two years after the publication of LHKSS, we review the observational
evidence for switched changes in magnetospheric states, both in
intermittent pulsars and in mode-changing pulsars, and discuss the
relationship between the two phenomena.

\section{Intermittent pulsars}
Many pulsars show intermittency in their radio emission, although
usually the durations of the "ON" and "OFF" states are measured in
seconds to hundreds of seconds, timescales which are far too short to
permit the determination of any change in slowdown rate between the
states.  This is the phenomenon of pulse nulling which has been known
since shortly after the discovery of pulsars (\cite[Backer 1970]{bac70}).

 \begin{figure}[t]
 \begin{center}
  \includegraphics[scale=0.31, angle=-90]{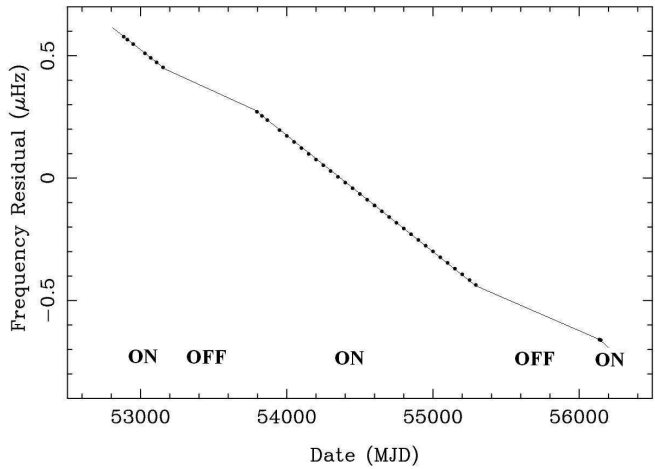} 
  \caption{The rotational frequency evolution of PSR~J1832+0029
 (\cite{llm+12}).} 
    \label{fig1}
 \end{center}
 \end{figure}

 \begin{figure}[t]
 \begin{center}
  \includegraphics[scale=0.31, angle=-90]{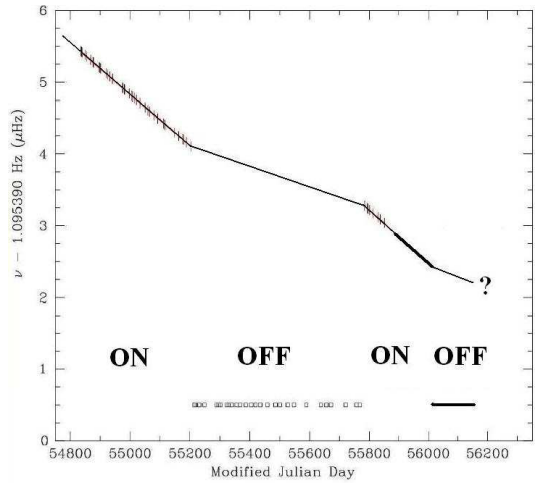} 
  \caption{The rotational frequency evolution of PSR~J1841$-$0500
 (\cite[Camilo et al. 2012]{crc+12} and Lyne, priv. comm.).} 
    \label{fig2}
 \end{center}
 \end{figure}

However, the intermittent pulsar B1931+24 is typically ON for 1 week
and OFF for about 1 month, permitting \cite[Kramer et
al. (2006)]{klo+06} to show that the ratio of ON- and OFF- slowdown
values $\dot{\nu}_{\rm ON}/\dot{\nu}_{\rm OFF}=1.5\pm0.1$, roughly
consistent with an absence of all magnetospheric currents during
the OFF phase, in accordance with the calculations of the braking
effects of magnetospheric currents by \cite[Goldreich \& Julian
(1969)]{gj69}.

Shortly after that publication, a second long-term intermittent
pulsar was discovered (PSR~J1832+0029) and reported to show similar
large changes in in slowdown rate ($\dot{\nu}_{\rm ON}/\dot{\nu}_{\rm
OFF}=1.7\pm0.1$; \cite[Kramer 2008]{kra08}, \cite[Lyne 2009]{lyn09},
\cite[Lorimer et al. 2012]{llm+12}).  Fig.~1 shows the measured values
of rotation rate during the 10 years since its discovery.  With rather
poor statistics, the lengths of the ON and OFF states are typically
many hundreds of days, compared with tens of days for B1931+24.

More recently, a third long-term intermittent object
(PSR~J1841$-$0500), also with timescales measured in hundreds of days,
has been reported by \cite[Camilo et al. (2012)]{crc+12}.  Even though
the statistics are also poor for this pulsar, it is clear that this
has an even greater slowdown rate ratio ($\dot{\nu}_{\rm
ON}/\dot{\nu}_{\rm OFF}=2.5\pm0.2$; see Fig.~2).

\section{Profile-switching pulsars}
LHKSS studied those pulsars in the Jodrell Bank timing database which
showed the largest amounts of timing noise, measured as the ratio of
maximum to minimum values of slowdown rate.  Seventeen examples are
shown in Fig.~3. Pulsars typically have peak-to-peak values of about
1\% of the mean, over a 4-orders-of-magnitude range of slowdown rates.
Individual pulsars may have a factor of 10 times more or less than
this.  LHKSS found that six of the 17 pulsars showed pulse-shape
changes which were correlated with these slowdown rate variations.
Subsequent studies (Lyne et al., in prep) have now shown that a
further four of these 17 also have significant pulse-shape variations
that are correlated with slowdown rate (PSRs B0919+06, B1642$-$03,
B1826$-$17 and B1903+07) as well as two others (PSRs B1740$-$03 and
B0105+65).

 \begin{figure}[hbt]
 \begin{center}
  \includegraphics[width=0.4\textwidth]{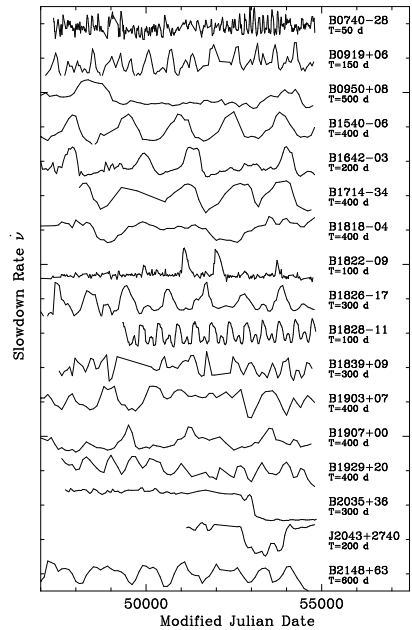} 
  \caption{The slowdown rate ($\dot{\nu}$) of 17 pulsars (from Lyne et
 al. 2010).} 
    \label{fig3}
 \end{center}
 \end{figure}

 \begin{figure}[hbt]
 \begin{center}
  \includegraphics[scale=0.35, angle=0]{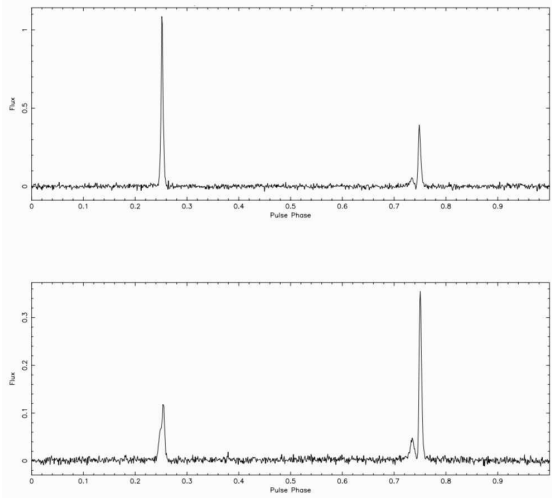} 
  \caption{The profiles of the ``normal'' (top) and ``abnormal''
(bottom) states of PSR~J2047+5029 (Janssen et al., in prep.).} 
    \label{fig4}
 \end{center}
 \end{figure}

A further striking example of correlated changes in pulse shape and
slowdown is displayed by PSR~J2047+5029, discovered at Westerbork in
the 8gr8 survey (\cite[Janssen et al. 2009]{jsb+09}).  At discovery,
the observed profile showed a main pulse and an interpulse with an
associated precursor having about 1/3 of the flux density of the main
pulse (Fig.~4 top).  However, when monitoring commenced at Jodrell
Bank, the main pulse had reduced in intensity by a factor of about 10,
so that it was then much weaker than the interpulse (Fig.~4 bottom).
The pulsar has since changed from this ``abnormal mode'' back to the
earlier ``normal'' mode.  These changes were accompanied by changes in
slowdown rate, which was larger when in the normal mode than in the
abnormal mode, again consistent with the notion that the particles
responsible for much of the normal-mode main-pulse radio flux density
were also responsible for the increase in braking.

In total there are now 16 pulsars with established synchronised
changes in the radio emission and the slowdown rate.  The properties
of these pulsars are summarised in Table~1, in decreasing order of the
magnitude of the timing noise, measured as the ratio of the maximum and
minimum slowdown rates.

\begin{table}
  \begin{center}
  \caption{Timing noise slowdown rate ratios ($\dot{\nu_1}/\dot{\nu_2}$) and emission changes in 16 pulsars.}
  \label{tab1}
  \scriptsize
  \begin{tabular}{|l|l|l|l|}\hline 
{\bf PULSAR} & {\bf $\dot{\nu}_1/\dot{\nu}_2$}~~ & {\bf Emission Change} & {\bf Reference} \\ \hline
J1841$-$0500~~ & 2.5  & Deep null & Camilo et al. (2012) \\
J1832+0029   & 1.7  & Deep null & Lorimer et al. (2012) \\
B1931+24     & 1.5  & Deep null & Kramer et al. (2012) \\ 
B2035+36     & 1.13 & 28\% change in W$_{\rm eq}$ & Lyne et al. (2010) \\
B1740$-$03   & 1.13 & 70\% change in component ratio & This paper \\
B0105+65     & 1.11 & 30\% change in W$_{\rm eq}$ & This paper \\
B1903+07     & 1.07 & 10\% change in W$_{\rm 10}$ & This paper \\
J2043+2740   & 1.06 & 100\% change in W$_{\rm 50}$ & Lyne et al. (2010) \\
B1822$-$09   & 1.033& 100\% change in precursor/interpulse~~ & Lyne et al. (2010) \\
J2047+5029   & 1.030& 90\% change in main pulse & Janssen et al. (in prep)~~ \\
B1642$-$03   & 1.025& 20\% change in cone/core & This paper \\
B1540$-$06   & 1.017& 12\% change in W$_{\rm 10}$ & Lyne et al. (2010) \\
B1828$-$11   & 1.007& 100\% change in W$_{\rm 10}$ & Lyne et al. (2010) \\
B1826$-$17   & 1.007& 10\% change in cone/core & This paper \\
B0919+06     & 1.007& 30\% change in component ratio & This paper \\
B0740$-$28   & 1.007& 20\% change in W$_{\rm 75}$ & Lyne et al. (2010) \\
\hline
  \end{tabular}
 \end{center}
\vspace{1mm}
 \scriptsize
\end{table}

\section{The nature of the switching}

The time sequences of slowdown rates shown in Fig.~2 are usually
bounded by well-defined maximum and minimum levels, each extreme level
being identifiable with a characteristic emission profile or flux
density.  As reported by LHKSS, each pulsar is usually seen to switch
abruptly between these extreme states.  The fact that the patterns in
Fig.~2 are generally smooth and do not display abrupt switching
behaviour was demonstrated to arise from changing statistical
properties of the mode-changing phenomenon, and the observed profile
shape parameter is determined by the proportion of time spent in the
two modes.  This is most clearly illustrated in Fig.~5 which shows a
number of 8-hour observations of PSR~B1828$-$11 chosen at different
phases of the 500-day oscillating pattern displayed by this pulsar in
Fig.~3.  We note that although pulse nulling and profile mode-changing
was first observed in 1970 (\cite[Backer 1970]{bac70}; \cite[Backer
1970a]{bac70a}), the following four decades have seen no study of the
stability of the statistics of nulling or mode-changing.


 \begin{figure}[bt]
 \begin{center}
  \includegraphics[scale=0.38, angle=-90]{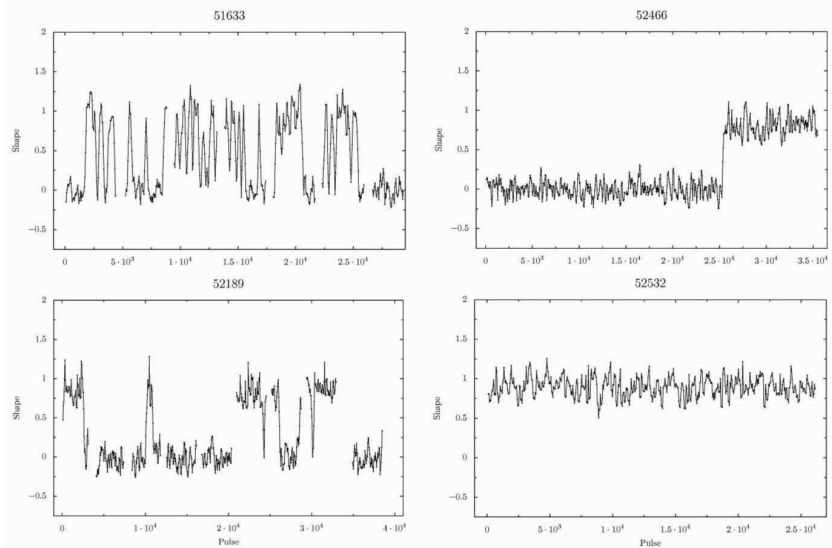} 
  \caption{Four 8-hour time-sequences of the profile state of
 PSR~B1828$-$11, taken at four different phases of the 500-day oscillation
 seen in Fig.~3 for this pulsar. Although the switching timescale may
 be short, the fraction of time spent in one state or the other
 changes slowly (Stairs et al. in prep.)} 
    \label{fig5}
 \end{center}
 \end{figure}

\section{The relationship between nulling and mode-changing}
The processes of nulling and mode-changing are similar in many
ways. Both are switched phenomena between (usually) two discrete
emission states.  They both have similar large ranges of timescales
and both have a major synchronisation with the spindown rate of the pulsar.
They are both understood in terms of changes in magnetospheric particle
currents.  The natural conclusion is that nulling is probably an extreme form of
mode-changing.  This view is supported by a few cases in which an
apparently nulling pulsar has been found to have low-level emission in
the null state.  One such example is PSR~B0826$-$34 shown in Fig.~6,
in which integration of the data during apparently ``null'' episodes
shows pulsed emission at a level of about 2\% of the un-nulled pulses.
 
Perhaps one telescope's nulling pulsar is a larger telescope's
mode-changing pulsar !

 \begin{figure}[hbt]
 \begin{center}
  \includegraphics[scale=0.35, angle=-90]{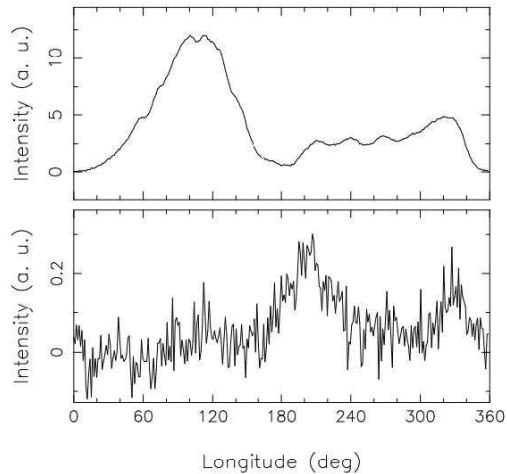} 
  \caption{The integrated ``normal'' (top) and ``abnormal'' (bottom) 
profiles of PSR~B0826$-$34 at 1374~MHz (from \cite[Esamdin et al. 2005]{elg+05}).} 
    \label{fig6}
 \end{center}
 \end{figure}

\section{Conclusions}
Pulsar magnetospheres switch between a small number of discrete
states, usually two, each of which corresponds to an apparently
quasi-stable magnetospheric configuration.  It seems that changes in
magnetospheric current flows between these states cause variations in
both the emission beam snd the slow-down rate. This is supported by
the general observation that the larger slowdown-rate is nearly
always associated with enhanced emission, particularly of the pulsar
``core'' emission.  I have no understanding of why there are these
discrete states, or of the origin of the multi-year quasi-periodicities that
modulate the statistical properties of the states.  Free-precession of
the neutron star and orbiting asteroids have been proposed, but any
links are obscure.

Finally, it must be emphasised that these phenomena are
widespread. The majority of pulsars of young and intermediate
characteristic age display detectable timing noise.  The studies of
profiles described here have mostly been of those pulsars which have
the largest fractional changes in slowdown rates and hence may be
expected to suffer the greatest magnetospheric changes and
corresponding variation in emission properties.  In fact, 12 of the 19
pulsars with largest timing noise show correlated emission
variations. Most of the remaining 7 have much poorer signal-to-noise
ratio, making the precise determination of pulse shape changes
challenging.  The changes expected in less timing-noisy pulsars are
likely to be much more subtle and a challenge to detect. At present,
there is no reason to doubt that all timing noise has its origin in
switched magnetospheric states.

\end{document}